\documentclass[aps,preprint]{revtex4}

\usepackage{graphicx}

\usepackage[utf8]{inputenc}

\begin{document}

\title{Shaping Magnetic Fields with Zero-Magnetic-Permeability Media}

\author{Alvaro Sanchez and Natanael Bort-Soldevila}

\affiliation{Departament de F\'isica, Universitat Aut\`onoma de Barcelona, 08193 Bellaterra, Barcelona, Catalonia, Spain}

\begin{abstract}

Some of the most important technological challenges of today's society, such as fusion reactors
for future clean unlimited energy or the next generation of medical imaging techniques, require
precise spatial shapes of strong magnetic fields. Achieving these high fields is
currently hindered by limitations such as large forces damaging the wires in coils or the saturation
of ferromagnets at high fields. Here we demonstrate a novel paradigm for creating magnetic landscapes. By enclosing magnetic sources within zero-magnetic-permeability
(ZMP) media, a set of novel properties is unveiled. The magnetic field shape directly results from the contour of the outer surface of
the ZMP enclosure, which allows the realization of basically any imaginable field landscape. Also, currents embedded in ZMP media can be fully magnetically isolated, which eliminates the forces in the wires, one of the main factors that currently impedes achieving very high magnetic fields. We confirm
these properties, rooted in fundamental laws of electromagnetism, by numerical simulations and by proof-of-principle experiments using conventional high-temperature superconductors as ZMP materials, which showcase the practical applicability of our ideas. The freedom in the design of magnetic fields provided by ZMP media enables to concentrate and homogenize magnetic fields with unprecedented precision, as needed in medical imaging techniques and particle-physics experiments, and to realize devices like perfect electromagnetic absorbers of mechanical vibrations.

\end{abstract}

\maketitle

\section{Introduction}

Magnetic fields power our society in motors, transformers, medical imaging techniques, and energy generators, from turbines in all kinds of electrical plants to future fusion reactors for clean unlimited energy \cite{coey,21cent}. Each of these technologies
require precise spatial distributions of magnetic fields, often with very high values \cite{coey,21cent}.
Magnetic fields are mainly created by current-carrying wires, forming cables, loops, and coils.  Fields emanate from wires always with the fixed shape of round field lines and strength decreasing inversely proportional to the distance.  Once created, the shape of the fields can be modified by ferromagnetic materials, which attract magnetic field lines \cite{coey,21cent}, and recently by ferromagnetic-based metamaterials as well \cite{wood,narayana,gomory,concentrator,hose,advmat}. However, all ferromagnetic materials become ineffective at fields larger than $\sim$2T, when they saturate \cite{coey}. Superconductors are another important tool to shape magnetic fields, in this case by excluding it in their interior. In particular, high-temperature superconductors (HTS), often made in the shape of thin tapes, bring important advantages over conventional superconductors, in terms of simpler cryogenics (owing to their higher critical temperature), increased stability, and lower power consumption \cite{ainslie}. However, the mechanical fragility of HTS tapes is severely compromising their use in many applications because of the enormous mechanical stresses they undergo when carrying high currents in the presence of strong fields (coil pressures in particle accelerator magnets, for example, can reach 100MPa, 1000 times atmospheric pressure \cite{kirby}). This is one of the most important obstacles hindering the future use of HTS in high-field applications \cite{marto, kirby,freidberg,choi,weijers}. A recent work \cite {larbalestier} clearly illustrates the current limitations for using HTS for ultra-high magnetic fields: a  superconducting-resistive coil beat the highest magnetic field
achieved until now but only for a very small margin (from 45T to 45.5T), due to the stresses appearing in the HTS tapes, which irreversibly damaged the coil after its use. 

All these problems set stringent limits to the set of available magnetic field shapes, and in particular in high field applications, when ferromagnetic materials become inefficient and superconducting tapes can be severely affected. Breakthroughs are needed for developing the future generation of magnets and for creating magnetic landscapes that can yield new applications and boost existing ones, and they should desirably incorporate ways to take full advantage of the potential of HTS.

Here we introduce a paradigm to create magnetic field landscapes based on directly modifying the field of the sources - wires, loops, and coils-, by surrounding them with bulk ZMP media, assumed as a linear material with a near-zero magnetic permeability $\mu$. A singular set of properties is unveiled, having direct applications in several key technologies, like MRI magnets, particle-accelerators or fusion magnets, and enabling the realization of novel devices like perfect electromagnetic vibration absorbers. We demonstrate that this control of static magnetic fields can be realized with currently available bulk HTS, which can endure huge magnetic fields \cite{Uglietti}, acting as ZMP media.

\section{Properties of ZMP media surrounding electrical currents}

Superconductors (or in general media with very small permeability, ZMP) can be considered in practice as perfect diamagnets, so that magnetic induction {\bf B}=0 in their interior (see section \ref{SC} for an analysis of HTSs as ideal ZMP media).  This property is currently used for shielding volumes from magnetic fields \cite{roadmap,denis}. Also, the field of a magnet or a coil fully enclosed inside a superconducting shell is shielded, i. e., the field does not leak outside the volume \cite{gomory}.

In contrast, the field of a current-carrying wire surrounded by ZMP media (Figs. 1 and 3) exits the media. In magnetostatics, from Amp\`ere's law, $\nabla \wedge {\bf  H}={\bf J}$, it results that magnetic fields of wires always escape from the superconducting enclosure (see Section \ref{physics} for the demonstrations of all the ZMP properties based on the magnetostatics laws).
Because $\nabla\cdot {\bf B}=0$, the perpendicular component of {\bf B} is zero at any border between air and ZMP media, so the field exiting the ZMP media is always tangent to the surface. The field of a wire enclosed by a ZMP medium is thus transformed into a field with the shape of the chosen outer surface of the ZMP medium. This property holds regardless of the position and shape of the hole in the ZMP medium and the location of the current in the hole. This is illustrated in Figs. 1(a)-(f) using finite-elements calculations for a current-carrying wire surrounded by a linear ZMP medium, both very long in the direction perpendicular to the depicted plane. Cases b-d give exactly the same external field distribution, that of the bare wire [Fig. 1(a)], whereas in Figures 1(e) and 1(f) the field changes, shaped by each external surface. 
This property of the field controlled by the source has a parallel in the control of electromagnetic waves by epsilon-near-zero (ENZ) media. The scattering response of a large arbitrarily shaped body with a single and arbitrarily located actuator embedded in the ENZ media was directly controlled by the media surface \cite{liberal,liberal_NatPhot}

All the properties of ZMP media enclosing currents hold not only in longitudinal but also in the practical case of closed current loops [axisymmetric case, Figs. 2(a)-(c) and 4(d)-(g), and Supplementary Figs. S1 and S2]. As an example with practical relevance,
in Fig. 2 we numerically demonstrate that the field of a coil composed of several current loops can be emulated by using only a single loop carrying the total current of the loops, enclosed by the adequately shaped ZMP media. Because bulk HTS can be realized with many shapes \cite{gozzelino}, a large variety of magnetic landscapes can be produced, even using as a source a single current loop.

We experimentally demonstrate these properties by enclosing a current-carrying wire with a Bi$_2$Sr$_2$Ca$2$Cu$_3$O$_{10+x}$ high-temperature superconducting hollow cylinder [Fig. 3(a)]. In Fig. 3(b) we show that the field measured in the superconductor exterior is always the same as if the cable was centered, independently of the actual location of the wire. In this way, any low-frequency vibration of the wire inside the superconducting cylinder would not yield an appreciable variation of the magnetic field outside. A perfect electromagnetic absorber of mechanical vibrations is thus experimentally realized. 

Even though our experiments shown in Fig. 3 are performed in static conditions, we can theoretically demonstrate that the field that exits a long superconducting cylinder that has a parallel current-carrying wire in its interior remains constant when the wire is moving. In Supppl. Fig. S6 and the Supplementary Video SV1, we use the module Rotating Machinery Magnetic
(rmm) of the finite-element solver Comsol to numerically simulate the case of a wire rotating at a frequency of 1Hz. There, it is clearly seen that the field of the moving wire, which is changing according to the wire movement, remains exactly unchanged in all exterior space when the wire is surrounded by the ZMP, an explicit demonstration of the perfect electromagnetic absorber of mechanical vibration behavior. 
There would be a physical limit for the perfect absorber property, given by the oscillation frequency of the wire. In the module used for the calculations of Supplementary Video SV1, it is assumed that magnetostatic Amp\`ere's law holds, which limits the validity of the results to frequencies at which displacement currents is negligible, i. e., the dimensions of the system are much larger than the wavelength of the resulting electromagnetic waves, as we discuss in Section VI.  We can therefore conclude that for all kinds of low-frequency vibrating wires, like those encountered in conventional electromagnetic machinery, the oscillations in the external magnetic field can be totally eliminated using ZMP media.

\section{Addition, cancellation and control of magnetic fields with ZMP media}

ZMP media can also be used to add, subtract or cancel currents at will. When several currents are surrounded by  ZMP media, the external field, shaped by the external ZMP surface, corresponds to the field created by the net current threading the media. The field of two parallel long wires each with current $+I$ [Fig. 4(a)] is transformed into the field of a centered single wire with current $+2I$ [Fig. 4(b)] by enclosing them within ZMP media. If the wires carry currents $+I$ and $-I$, respectively, then the external field is exactly canceled [Fig. 4(c)]. An important application is demonstrated in [Fig. 4(c)]: we obtain the properties of a coaxial cable (one that carries two opposite currents with zero field outside),  but having complete freedom for choosing the location of each current. 

Figs. 4(d)-(g) show other examples of the rich phenomenology of current additions by ZMP media, now for current loops. The field of the current loop [Fig. 4(d)] surrounded by a solid ZMP piece [Fig. 4(e)], is confined to the void where current sits. However, if the topology changes to a hollow disk, now field exits the media and actually concentrates in the central volume enclosed by the disk [Fig. 4(f)]. If in this situation a second current with opposite sign is added in a concentric hole in the ZMP medium, then net current cancellation makes the field outside zero again [Fig. 4(g)]. The latter situation deserves special attention: the field of a current loop with current $+I$ is exactly canceled by a different loop, with arbitrary radius, carrying a current $-I$.

The addition properties of ZMP media, including the fact that the exterior magnetic field is additive and independent of the position of the holes or the wires in the holes, is also reminiscent of the properties of ENZ media for electromagnetic waves: dielectric rods immersed in a 2D  ENZ medium modify the effective permeability of the entire structure in an additive manner, independently of the position of the rods \cite{liberal,liberal_NatPhot}.

We experimentally demonstrate the addition properties [Fig. 5(b)] by measuring the field created by two wires carrying different currents surrounded by a YBa$_2$Cu$_3$O$_{7-x}$ HTS cylinder with two holes carved along its axis [Fig. 5(a)]. The cylinder is actually made of several stacked disks, showing that ZMP media can be well emulated even by multiple HTS pieces (see Supplementary Fig. S5).   

\section{Removing the forces in the wires of electromagnets}

Two other important features of currents enclosed in ZMP media lead to further applications. Firstly, the field that a current in one hole creates in another hole in the same ZMP media is exactly zero, as physically explained in Section VI and  experimentally demonstrated in Fig. 6. There is no magnetic interaction between currents contained in different holes within ZMP media (different from the case of both currents in the same hole, in which they interact normally).
Secondly, the field created in the vicinity of a current sitting in a hole in a ZMP medium is exactly that of the bare current as long as it is centered and the hole surface has the shape of a field line (which is a circumference for longitudinal geometry, like Figs. 1(c), 1(e) and 1(f), and some elliptical-like shapes for axisymmetric cases, see Supplementary Fig. S1). The two properties imply that by adequately placing the current(s) in the hole(s) in the ZMP medium one can completely isolate magnetically these currents: from other currents in the ZMP medium, from externally applied fields, and also from the ZMP medium itself. Being magnetically isolated, currents cannot interact with anything, and thus do not experience any force. In contrast to this situation, when the current wire is not properly centered or when the shape of the hole in the ZMP media does not correspond to a field line, then the field around the wire is distorted. A restoring force occurs in this case, that tends to bring the wire back to the equilibrium position, at which there is no force. This situation is reminiscent of the case of a magnet levitating on top of a symmetric superconductor, e. .g , a superconducting ring \cite{reviewCSM}. In both cases, there is a repulsive force between the magnetic source- magnet or wire- and the superconductor, and a restoring force towards the equilibrium position.

The strategy of surrounding superconducting wires or tapes with ZMP materials solves one of the main problems in current high-field magnets \cite{kirby,freidberg,choi,weijers,marto,natphys}, since it would allow YBaCuO tapes to be employed in the highest-field environments. The superconducting tapes in high-field electromagnets could be placed in different holes within the ZMP media.  
In this way the tapes would be isolated from any magnetic field and thus would be spared from the large magnetic forces, which would be transferred to the surrounding bulk ZMP material, that can be realized with bulk superconducting materials. Bulk HTS with robust mechanical properties have shown to withstand large forces in superconductor magnets \cite{tomita,roadmap}.
The magnetic isolation provided by ZMP media also would protect HTS tapes from the induced shielding currents on their wide face, which deteriorate the field quality of magnets \cite{amemiya}.

\section{Applications of ZMP materials to improve high-field magnets}
\label{magnets}

The set of properties provided by ZMP media may have an important impact on fusion magnets, particle accelerators, MRI medical imaging techniques, and other high-magnetic-field technologies. 

We show in Fig. 7 how selected ZMP enclosures can transform the field of a simple coil into shapes with technological interest, including ways to concentrate magnetic fields in unprecendented ways. The usual strategy of using the gap between two ferromagnets breaks down at high field because of their saturation at around 2T \cite{coey}. Instead, a large field can be concentrated using constrictions made with ZMP media consisting of HTS -for example, 
YBa$_2$Cu$_3$O$_{7-x}$ or Bi$_2$Sr$_2$Ca$2$Cu$_3$O$_{10+x}$, which can be active under much more intense fields, because of HTS huge critical fields  \cite{grisso,godeke}.  In Fig. 7(f) we illustrate how a 'magnetic bottle', a volume limited by two very intense field regions (magnetic 'mirrors'), relevant for linear fusion reactors \cite{linearfusion1,linearfusion2,natphys}, can be created from simply a single coil. The freedom for magnetic design offered by the addition and/or activation of currents embedded in ZMP media is vast.

As another relevant example, we show in Fig. 8 how in a hybrid resistive-superconducting magnet like the one that broke the magnetic field record on earth [Fig. 8(a)] \cite{larbalestier}, by surrounding the central HTS coil with an HTS enclosure [Fig. 8(b)] one can completely remove the forces that damaged the wires, while achieving a similar magnetic field value in the central region. 
By adequately shaping the HTS enclosure one can even largely improve the field strength and homogeneity  in the gap or change the field shape at will [Fig. 8(c)]. 
The property of enhancing the field homogeneity [as shown in Figs.  2(b) and 8(c)] is actually a very relevant one for some cutting-edge recent experiments. In magnetic resonance imaging, MRI, poor main magnetic field homogeneity leads to artifacts and signal losses, and therefore becomes one of the main obstacles for achieving good-quality images for medical diagnostics \cite{chen_mri}.
Another relevant example is the Muon g-2 experiment in Fermilab, where very precise measurements of the muon magnetic moment may show discrepancies with theoretical predictions of the standard model \cite{albahri}. 
Because the muon precession frequency is proportional to the magnetic field, it is required that the average magnetic  field experienced by the muons remain stable on the scale of parts per million throughout the experiment. A very homogeneous field is required to minimize the uncertainty of the magnetic-field maps caused by any nonuniformities in the muon 
distribution \cite{albahri}. Our results can provide a tool for improving the magnetic field homogeneity in these applications.  By enclosing the field-generating coils with ZMP, which can be realized by bulk HTS with very homogeneous composition and very flat surfaces [see the  example of the field profile in blue in Fig. 8(d)], the field homogeneity can be significantly improved in future experiments and devices.

\section{Demonstration of the properties of ZMP media by magnetostatic theory}
\label{physics}

We now show that all the properties of zero-magnetic-permeability (ZMP) media enclosing currents experimentally demonstrated and corroborated by numerical simulations in the previous sections can be derived from the basic equations of magnetostatics, which we briefly recall first.

When dealing with static conditions, electric and magnetic field decouple \cite{wood}. If one is only interested in the magnetic properties, physics is described by the two magnetostatic Maxwell equations $
\mathbf{\nabla} \times \mathbf{H}=\mathbf{J}$, and $\mathbf{\nabla}\cdot \mathbf{B}=0$,
where $\mathbf{H}$ is the magnetic field, $\mathbf{B}$ is the magnetic induction, and $\mathbf{J}$ is the free current density. For simplicity, in this work we refer to {\bf B} as magnetic field. 
Two boundary conditions result from these equations, when applied at any boundary between different magnetic media 1 and 2,
\begin{equation}
\mathbf{n} \times (\mathbf{H_1}-\mathbf{H_2})=\mathbf{K}, \label{boundH}
\end{equation}
\begin{equation}
\mathbf{n} \cdot (\mathbf{B_1}-\mathbf{B_2})=0, \label{boundB}
\end{equation}
where $\mathbf{K}$ is the free surface current density and $\mathbf{n}$ is a unit vector perpendicular to the interface. That is, the component of {\bf B} perpendicular to the surface is always continuous and the component of {\bf H} parallel to the surface is only continuous when there is no surface current.

The differential equation $
\mathbf{\nabla} \times \mathbf{H}=\mathbf{J}$ can be expressed in integral form, known as Ampere's law, as
\begin{equation}
\oint_C{\bf H}\cdot{\rm d}{\bf l}=I, \label{intH}
\end{equation}
where $I$ is the free current threading any surface limited by the one-dimensional line $C$.

{\bf B} can be expressed as the curl of a potential vector {\bf A}, $
\mathbf{B}=\mathbf{\nabla}\times \mathbf{A}$.
In the longitudinal case of translation symmetry, {\bf A} has only $z$ component, whereas in the axisymmetric case of rotational symmetry, {\bf A} has only the angular $\phi$ component. That is, in the two geometries that we study in this work, longitudinal and axisymmetric, {\bf A} can be described as scalar fields.
Assuming the Coulomb gauge, $\mathbf{\nabla}\cdot \mathbf{A}=0$, 
{\bf A} fulfills in vacuum Laplace's equation
$
\nabla^2 {\bf A}=0,
$
where it is understood that the equation is valid for each component of {\bf A}.
There is an analogy between the vector potential, which can be represented as a scalar field in the two geometries studied, and the scalar potential in electrostatics. The condition for constant scalar potential in a conductor in electrostatics is equivalent to  constant value of $A_z$ or $A_\phi\rho$, for the longitudinal and axisymmetric cases, respectively, in ZMP media. 

Based on the above basic formulas of magnetostatic theory, in the following we explain the properties described in the previous sections. We assume that ZMP media is a linear material with zero permeability; below we discuss how good this approximation is for actual high-temperature superconductors.

The property that the field created by a current surrounded by ZMP media has the shape of the external surface of the ZMP (illustrated in Figs. 1, 2, and 3)
results directly from the combination of Ampere's law, Eq. \ref{intH}, and the boundary condition for {\bf B}, Eq. (\ref{boundB}), at the surface of the ZMP media. Because of the boundary condition, the normal component of {\bf B} is zero, so that {\bf B}, which is different from zero because of Ampere's law, is tangent at the surface. It is thus demonstrated that this property holds regardless of the 
the position of the hole in the ZMP medium, the position of the wire in
the hole, and the shape of the hole.

The property of addition and subtraction of currents shown in Figs. 4 and 5 also results directly from Ampere's law and the arguments of the preceding paragraph. The line integral of {\bf B} just outside the outer surface of a ZMP media enclosing an arbitrary number of currents is directly proportional to the net current threading the area closed by the line, thus demonstrating the addition/subtraction results.

The fact
that a current in one hole creates a zero field in another hole enclosed in a ZMP medium, experimentally demonstrated in Fig. 6, also results from fundamental magnetostatics laws. We state the problem in terms of the magnetic potential vector.
The inner surface of the hole is an equipotential surface of {\bf A} (i. e., ${\bf A}_z$ for longitudinal cases and $A_\phi\rho$ for axisymmetric cases), because {\bf B} is zero in the ZMP medium. Since there is no current within the cavity and we assume magnetostatic conditions, the potential inside the cavity must satisfy Laplace's equation $
\nabla^2 {\bf A}=0,
$ and, therefore, it cannot have a local extrema (maximum or minimum) within the hole. Thus, the vector magnetic potential within the hole must be constant, and therefore {\bf B} in the hole must be zero. This property has also an analogous counterpart in electrostatics (with the electrostatic potential fulfilling Laplace's equation in this case), and is used to construct electrical shielding devices like Faraday's cages.

Finally, the property that the field created in the
vicinity of a current sitting in a hole in a ZMP medium
is exactly that of the bare current as long as it is centered and the hole surface has the shape of a field line
[which is a circumference for longitudinal geometry, like
figs. 1(c), 1(e) and 1(f), and some elliptical-like shapes for
axisymmetric cases, see Supplementary Fig. S1], can be derived from the uniqueness theorem of magnetostatics. Because this solution fulfills Laplace equation 
$\nabla^2 {\bf A}=0$ in all the hole and the boundary conditions [field {\bf B} tangent to the surface,  from Eq. (\ref{boundB})],  then it is the correct solution.
In contrast, when the shape of the hole is not that of a field line or the wire is not centered in the corresponding center of the field line, the field around the current will be modified with respect to the field of the bare current. This is because when approaching the border to the ZMP media, the field lines will change their shape in order to satisfy the boundary condition for {\bf B}, Eq. (\ref{boundB}), and thus be tangent to the ZMP surface. This will make the field around the wire to be modified with respect to the case when there is no ZMP material, thus yielding a magnetic interaction between the wire and the ZMP material in this case.

All the formulas in this section are exactly valid in the magnetostatic limit, but also approximately valid for low-frequency AC magnetic fields. As long as displacement current ($\partial {\bf D}/\partial{t}$, where {\bf D} is the displacement vector field) can be neglected, the magnetostatics solutions can be applied. The condition can be alternatively stated as the spatial variation of fields being much larger than the dimensions of the materials involved. 
This sets a limit in the frequency of a few hundredths 
of kilohertzs if dealing with materials on the order of tens of centimeters. Previous works have demonstrated experimentally the validity of these assumptions for low-frequency magnetic fields \cite{accloak,advmat}.

\section{Adequacy of high-temperature superconductors as ZMP materials}
\label{SC}

We have assumed in this work that  ZMP materials are linear media with a magnetic permeability that has a very small, 'near-zero', value. In this subsection, we argue that superconductors, and in particular, high-temperature superconductors (HTS) can fulfill in some regimes the properties for being regarded as ZMP materials.

It is well known that superconductors can be described, in a simplified scheme, as perfectly diamagnetic materials since {\bf B}=0 in their interior. However, in practice, the situation is more complex. In the following we will be referring to type-II superconductors, because type-I superconductors have less practical importance and are not used in high-current applications \cite{tinkham}. 
All superconductors employed in actual applications like particle accelerators, high-field magnets or MRI equipment, typically Nb alloys, are type II superconductors, as so they are all HTS materials \cite{tinkham}. 
The modelling approach
used in the paper ($\mu$ tending to zero) is strictly valid only below the lower critical field ($B_{\rm c1}$), which
is very small for HTS.
Above $B_{\rm c1}$, in the mixed state, type-II superconductors are non-linear and show hysteresis when an external field or voltage is applied to them and then removed, because of the vortices that appear \cite{tinkham,reviewCSM}. However, in the Nb alloys and the HTS typically used in strong-field applications there is a very strong pinning of the vortices in defects in the superconductor. In these situations, superconductors exhibit a very large value of the critical-current density $J_{\rm C}$, and are well described by the critical-state model \cite{reviewCSM,sirois}. Shielding currents appearing when an external magnetic field is applied are mostly confined to the surface of the material. Therefore, in strong-pinning superconductors like HTS or Nb alloys, the bulk of the superconductor is protected from the external fiels by these high surface currents, and the approximation that {\bf B} inside the bulk is very small (i.e. the permeability $\mu$ is near zero) is valid in most practical cases. This has been experimentally demonstrated in many occasions, for low applied magnetic-field values. Magnetic metamaterials are good examples of that: magnetic cloaks that conceal magnetic fields \cite{gomory}, hoses that transfer magnetic fields \cite{hose} and concentrators of magnetic energy \cite{concentrator,exp_concentrator} have been theoretically derived assuming $\mu$ near-zero values in the superconducting parts, and have been experimentally demonstrated using different HTS materials, showing an excellent agreement with theory. 
In Figs. 3, 5, and 6 we have experimentally demonstrated the good adequacy of HTS as ZMP media in the cases of two different HTS materials, YBa$_2$Cu$_3$O$_{7-x}$ or Bi$_2$Sr$_2$Ca$2$Cu$_3$O$_{10+x}$, and even when they are made of adjoining pieces instead of a compact bulk material.

In spite of the large values of the critical-current density of HTS, some degradation of the superconducting properties could be expected when dealing with large magnetic fields, which would result in values of the effective permeability $\mu$ departing from zero. We have roughly estimate these effects by recalculating the configuration of Fig. 8c, assuming now two different scenarios with weakened-$\mu$ values. In Supplementary Figure S3 we consider that the ZMP has decreased values of $\mu$, homogeneously in all its bulk. In Supplementary Figure S4, the
drop in the $\mu$ values is assumed to occur in a surface layer, which would correspond to a material with a large critical-current value that confines the penetration of magnetic field to the region close to the surfaces of the superconductor \cite{mag_navau,reviewCSM}. 
In both cases, we see that in general the large values of concentration of magnetic field and its spatial homogeneity are somehow preserved even when the permeability of the superconducting material is reduced by more than one order of magnitude.
However, in a realistic scenario the depletion of the superconducting properties would set some limit of applicability of our ideas. 
The case in Supplementary Figure S4 can be a first-order representation of the effect of the penetration of currents in type-II superconductors at applied fields $B_{\rm a}$, whose effective depth is controlled by the critical current density, $\lambda_{\rm eff} = B_{\rm a}/(\mu_0 J_{\rm C})$  \cite{reviewCSM}. Assuming typical parameters for high-temperature superconductors, e. g. $J_{\rm C}\sim 10^{9}$A/m$^2$ \cite{sunwong}, a field of 10T would correspond to a current, and field, penetration of about 1cm. These simple calculations indicate that HTS will rather adequately play the role of ZMP for fields up to few Tesla, whereas their ideal performance can become compromised when dealing with larger applied fields or very small structures.

Finally, the fact that HTS has to be refrigerated should not be a serious inconvenience for applying our ideas. We have verified the good properties of YBa$_2$Cu$_3$O$_{7-x}$ or Bi$_2$Sr$_2$Ca$2$Cu$_3$O$_{10+x}$ materials in our proof-of-concept experiments by simply submerging the superconductors in liquid nitrogen. In most, if not all, of the high-field applications we mention in our work, from particular accelerators to fusion magnets to MRI devices, cryogenic environments at few kelvin temperatures are routinely used. It is also important to remark that all our experiments in this work were performed in a zero-field-cooled regime \cite{reviewCSM}, i. e., by cooling the superconductor in zero field, and then applying the magnetic field or setting the current in the wire(s). In general, in a field-cooled situation, in which the superconductor is first exposed to a magnetic field and then cooled, the magnetic flux that was present when cooled down is kept in the superconductor, with the possible risk of departing from the $\mu$ near-zero property. However, a simple solution to ensure that HTS function as ZMP media under any field-cooled regime is to break the superconducting path. By creating a slit in the ZMP or make it by two adjoining pieces instead of a single one, the basic diamagnetic property equivalent to $\mu\sim0$ would be recovered. This strategy was discussed and experimentally validated in the development of a magnetic hose \cite{hose}.
In this regard, we have demonstrated that HTS made of pieces work very well as ZMP media: the YBa$_2$Cu$_3$O$_{7-x}$ tube used for the results in Figs. 5 and 6 was actually made of ten disk-pieces stacked on top of the other; it reproduced very well the behavior theoretically predicted for a solid piece.

\section{Conclusions}

We have introduced a toolbox for creating magnetic fields by directly modifying the field of sources, instead of the fields once they are created, extending the current limits of electromagnetism. Fields of currents enclosed by zero-magnetic-permeability (ZMP) media can be added, canceled, controlled by other currents, and shaped at will. Some properties are reminiscent of those for controlling electromagnetic waves with ENZ materials; whereas ENZ media are very difficult to realize \cite{kinsey}, ZMP media for static fields are readily available using high-temperature superconductors. We have experimentally demonstrated these features using two types of superconductors as ZMP media, one even made of stacked pieces, which indicates the practical applicability of our ideas. Our results open novel routes for solving some of the main current problems for achieving high magnetic fields, like eliminating the forces in the wires. They also enable achieving features like unprecedented spatial homogeneity and concentration of magnetic fields, essential for applications like particle-physics experiments and fusion magnets, or creating the field of a coil from a single current loop, as well as allowing devices such as perfect electromagnetic vibration absorbers. 

\section*{SUPPLEMENTARY MATERIAL}

See supplementary material for methods, supplementary figures S1-S6, and description of supplementary video SV1.

\section*{Acknowledgements}

We thank C. Navau, A. Palau, M. Trupke, and I. Liberal for discussions, J. Plechacek for help preparing the superconducting samples, and R. Mach and S. Laut for initial assistance. A. S. acknowledges funding from ICREA Academia, Generalitat de Catalunya.

\section*{Availability of Data}
The data that support the findings of this study are available from the corresponding author upon reasonable request.

(*) Corresponding author: alvar.sanchez@uab.cat.

\newpage

\begin{figure*}[t]
	\centering
		\includegraphics[width=1.0\textwidth]{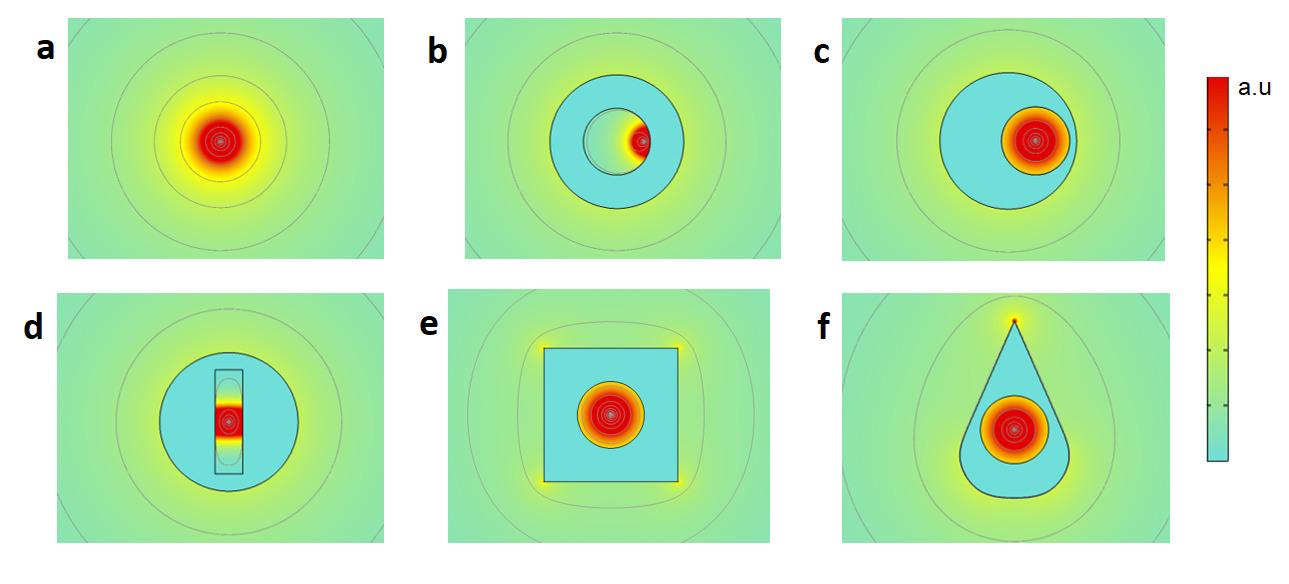}
	\caption{
 Magnetic field {\bf B} modulus for (a) a bare straight current, and (b-f) differently shaped linear ZMP media surrounding a straight current, both very long in the direction perpendicular to the figures. The magnetic field in the exterior of the ZMP material depends only on the shape of the outer surface of the ZMP material and the current of the wire, whereas the magnetic field inside the hole depends on the shape of the hole, on the wire position in the hole and on its current (analogous results for the axisymmetric case are shown in Supplementary Fig. S1). 
Magnetic field lines are tangential to the surface in all air-ZMP material boundaries. The field in the hole is the same as that of the bare current in (a), indicating complete magnetic isolation, when the hole is circular and the wire is centered in the hole [cases (c), (e), and (f)].  }
\end{figure*}

\newpage

\begin{figure*}[t]
	\centering
		\includegraphics[width=1.0\textwidth]{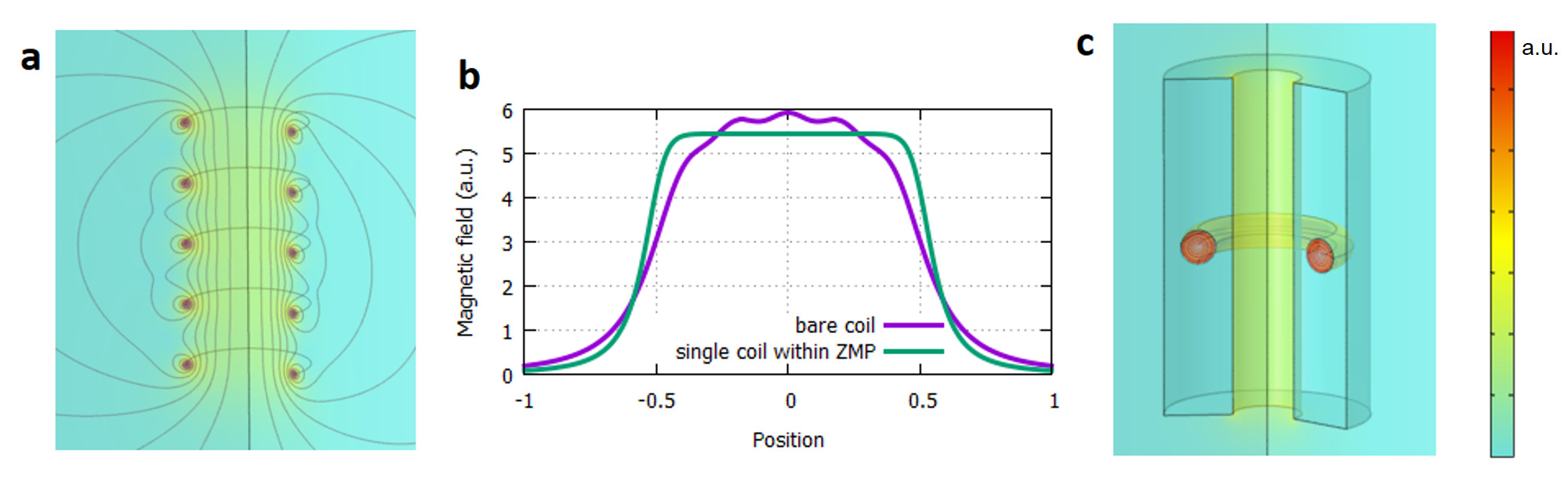}
	\caption{
(b) Comparison of the magnetic field {\bf B} at the vertical axis of a five-loop coil (a), in purple, and a single current loop (with five times the current of a loop in the coil) surrounded with ZMP material, (c), in green. A similar -even more homogeneous- field profile is obtained with the single loop in the ZMP medium.  }
\end{figure*}

\newpage

\begin{figure*}[t]
	\centering
		\includegraphics[width=0.7\textwidth]{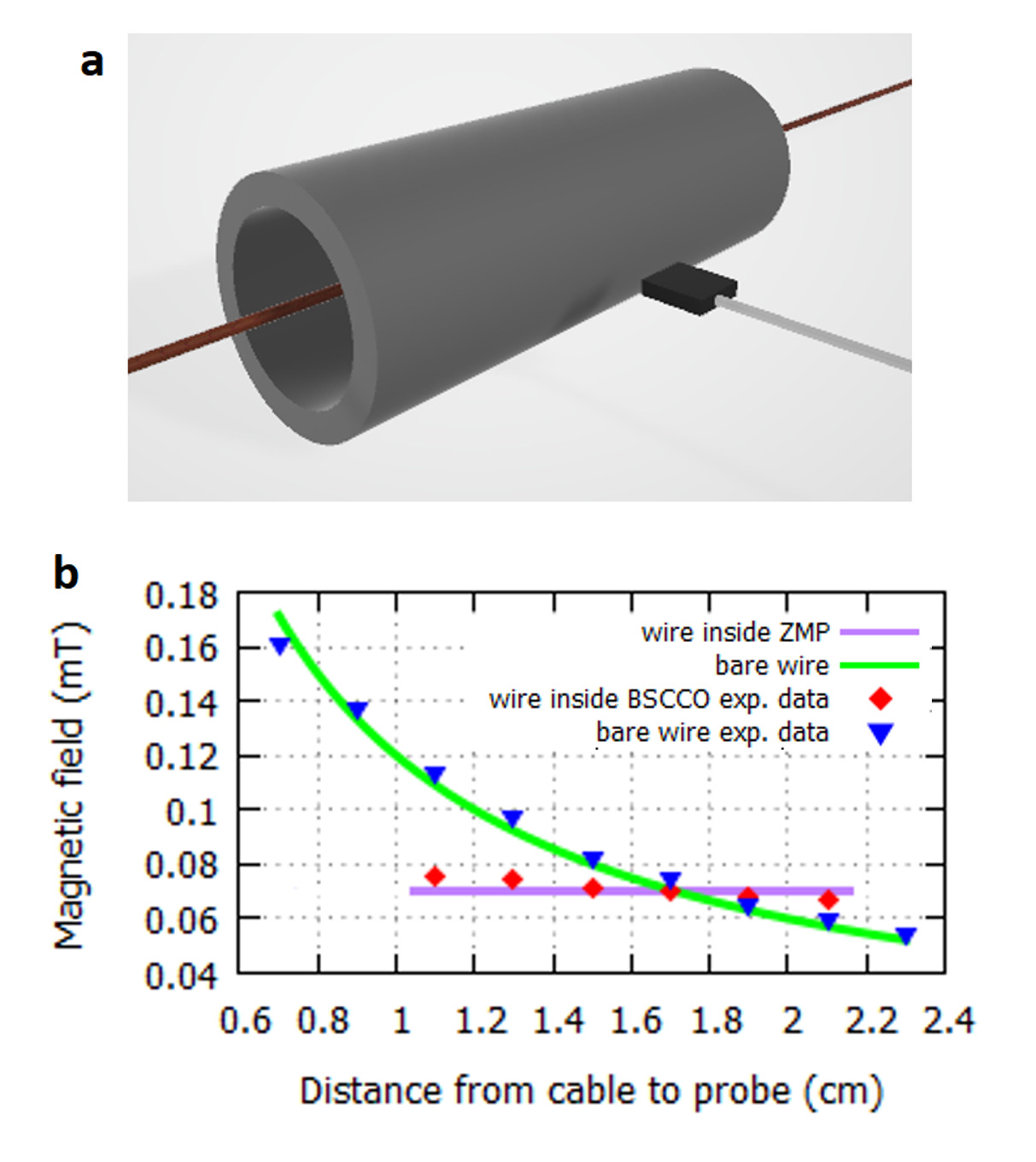}
	\caption{
	 {(a)} Sketch of the experimental setup of the cylindrical BiSrCaCuO superconducting material, the copper current wire and the Hall probe which measures the magnetic field {\bf B} (see Methods for experimental details). {(b)} Experimental results of the magnetic field created by a current wire carrying 6A placed at different distances from the probe, inside the BiSrCaCuO cylinder (red squares) and without the BiSrCaCuO material (blue triangles). The green line is the theoretical dependence of {\bf B} created by the wire as a function of distance from the probe. Error bars are smaller than the symbol size. {\bf B} remains constant for all the wire positions inside the ZMP material, maintaining the {\bf B} value as if the wire were always located at the cylinder axis (purple curve); a small experimental deviation can be attributed to the short length of the SC tube.}
\end{figure*}

\newpage

\begin{figure*}[t]
	\centering
		\includegraphics[width=1.0\textwidth]{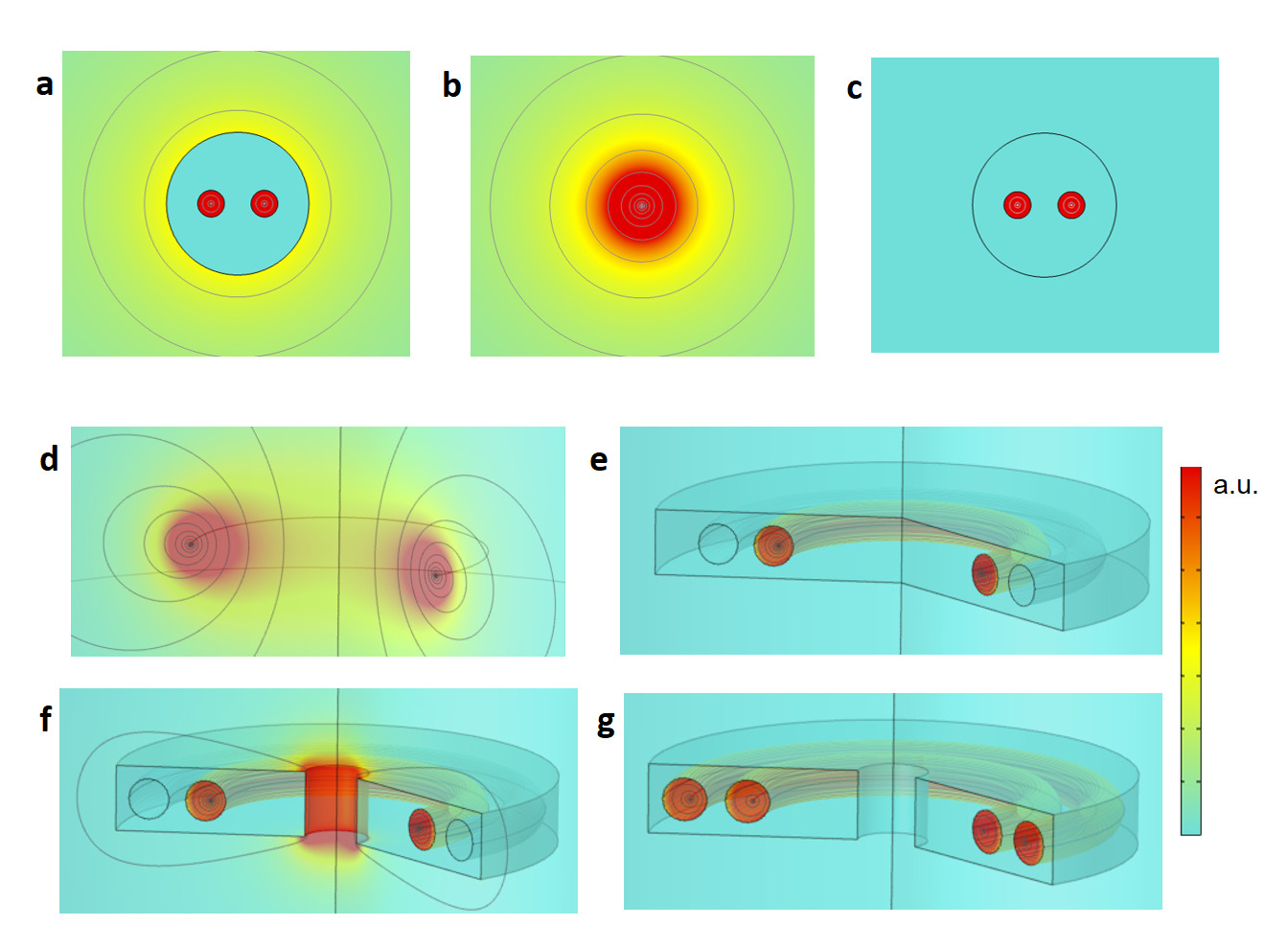}
	\caption{
 Magnetic field {\bf B} modulus for {(a)} two wires each with current $I$ with same sign in two different holes in a cylindrical ZMP medium, {(b)} a single wire with current $2I$ and {(c)} same as {(a)} but currents with opposite signs. Currents and media are very long in the direction perpendicular to the figures. {\bf B} in the media exterior in {(a)} is exactly as in {(b)}, that is field outside is the same as if there is a single centered wire with 2$I$, 
whereas in {(c)} the different sign of currents yields zero exterior field. Field always corresponds to the addition of the enclosed currents (analogous results for the axisymmetric case are shown in Supplementary Fig. S2). The field of a current loop {(d)} becomes basically concentrated in the central region when enclosed by a ZMP medium with a central hole {(f)}, in contrast with {(e)} the situation when the enclosing ZMP media is simply connected, for which {\bf B} is zero outside. {(g)} When there are two currents with opposite signs enclosed by the non-simply-connected ZMP media, {\bf B} is also zero outside. }
\end{figure*}

\newpage

\begin{figure*}[t]
	\centering
		\includegraphics[width=0.7
	\textwidth]{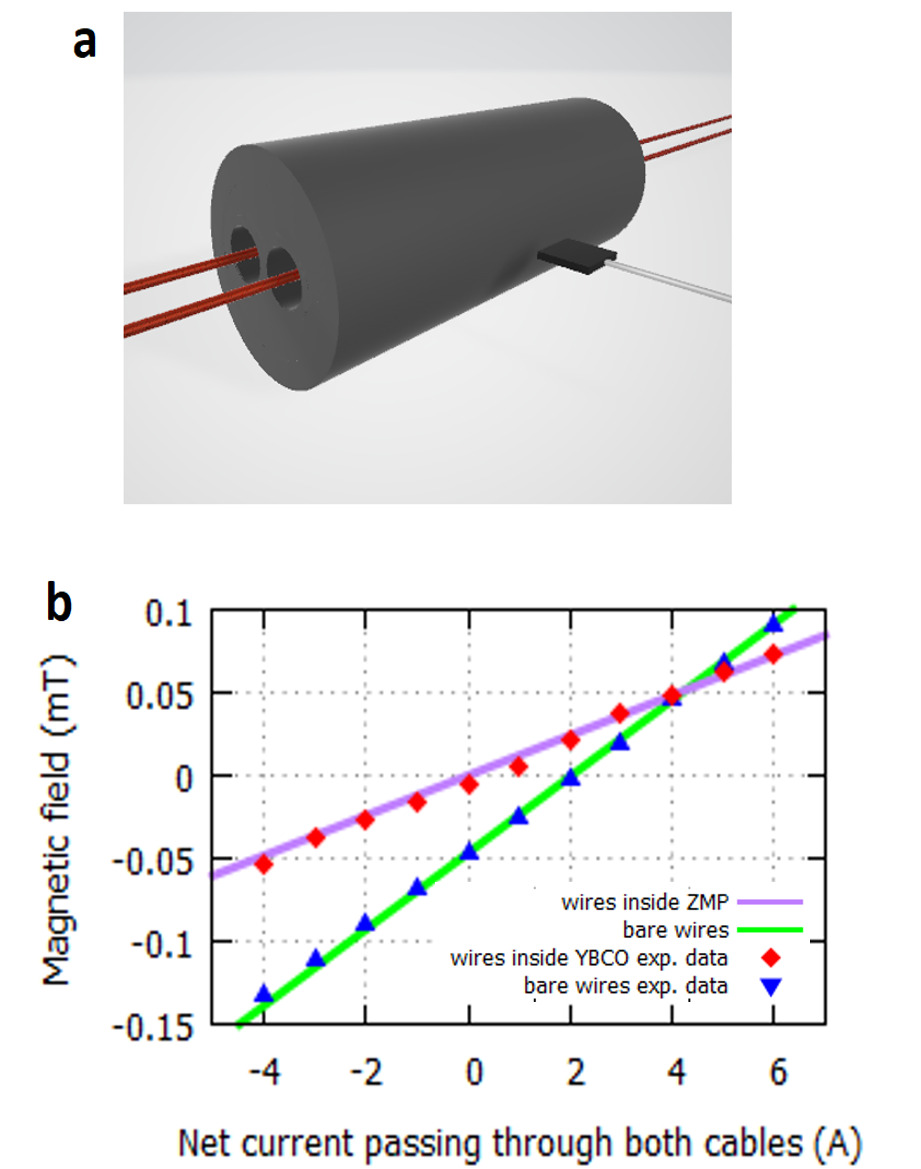}
	\caption{
(a) Sketch of the experiments for demostrating addition of currents, with a YBaCuO superconducting tube (made of 10 stacked disks) surrounding two wires (see Methods for experimental details).  (b) Measurements of {\bf B} at the probe position with constant current of +3A in one wire and a varying current (from -7A to +3A, in steps of 1A) in the second one (the closest to the probe), as function of the net current passing through both wires. Error bars are smaller than the symbol size. When the ZMP material is surrounding the wires [as sketched in (a)], red symbols, field values coincide with the theoretical field created by a centered wire carrying the net total current (purple line). In contrast, when there is no ZMP material, blue symbols, the measured {\bf B} coincides simply with the addition of the field of the two wires (green line). }
\end{figure*}

\newpage

\begin{figure}[ht]
\centering
\includegraphics[scale=0.45]{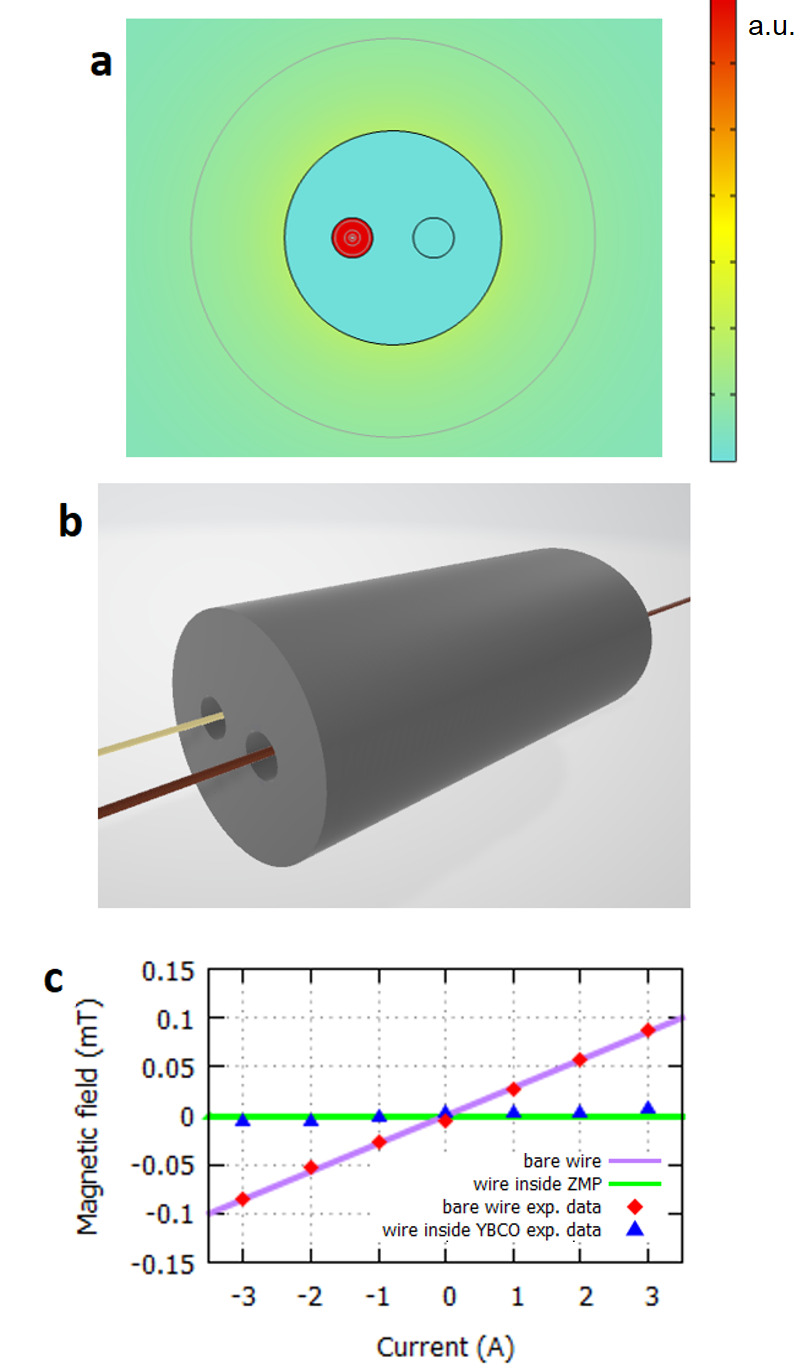}
\caption{
(a) Magnetic field {\bf B} modulus for a ZMP media with two holes, one with a current wire and a second one empty. {(b)} Sketch of the experimental setup, with a YBaCuO superconducting tube with two holes (grey), one with a current wire (brown) and the other with a Hall probe (white) measuring the magnetic field. {(c)} Experimental values of {\bf B} inside the hole with the Hall probe, for different currents in the first hole (blue triangles), and the same when the ZMP material is removed (red diamonds). Error bars are smaller than the symbol size. It is experimentally demonstrated that the field in the empty hole surrounded by ZMP material is zero for all values of current.
} 
\end{figure}

\newpage

\begin{figure}[ht]
\centering
\includegraphics[scale=0.4]{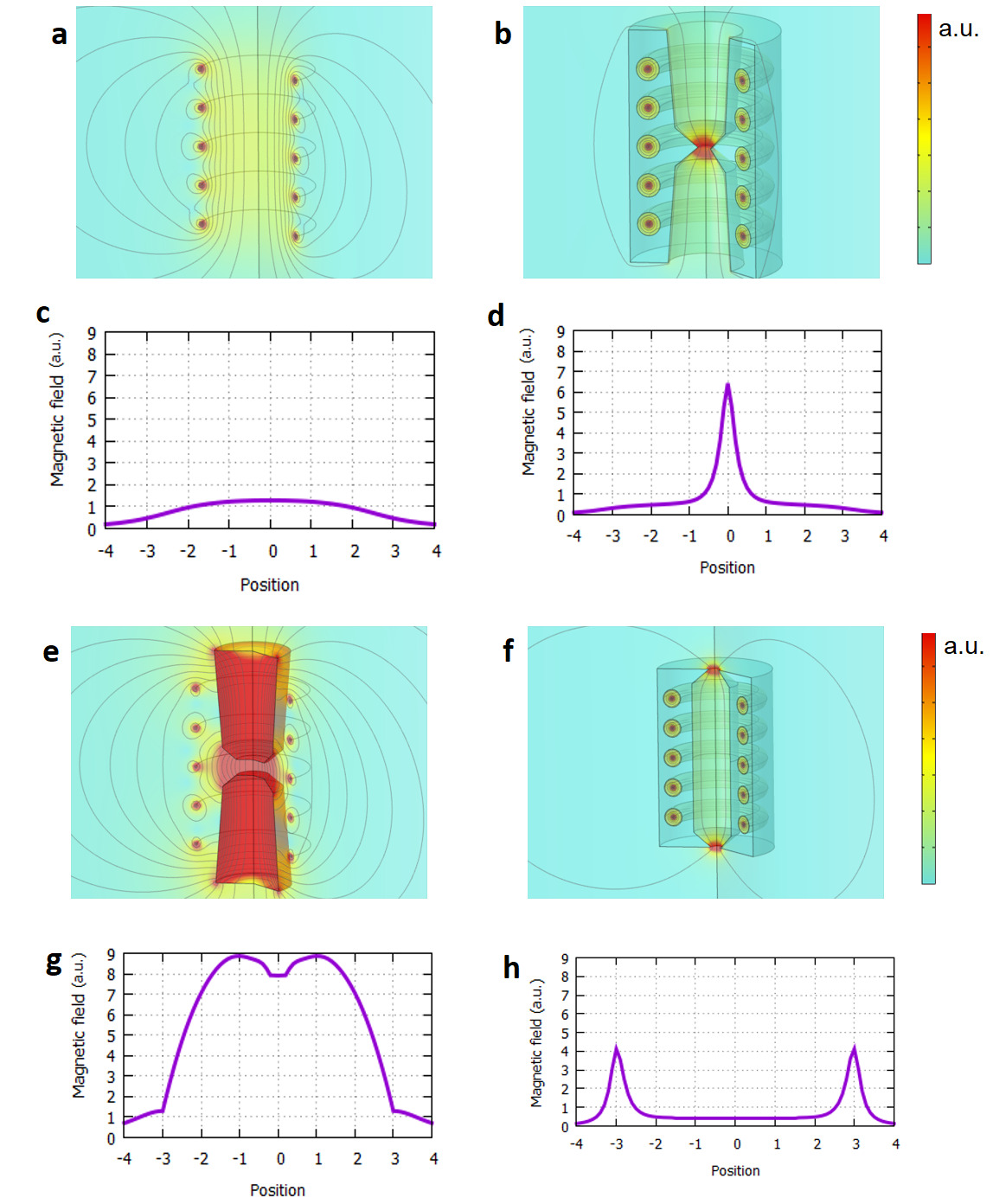}
\caption{
Magnetic field {\bf B} modulus for (a) a coil, and the same coil surrounded by different materials. In {(b)} and {(f)} ZMP media is used to concentrate the field in one and two points, respectively. The configuration in {(f)} corresponds to a 'magnetic bottle', a volume limited by two very intense field regions (magnetic 'mirrors'), relevant for
linear fusion reactors. In {e)} we show an alternative - conventional- way to concentrate the field of the coil in one point, as in {(b)}, using in this case a ferromagnetic material. However,  the configuration in {(e)} would not be valid at high fields, because of the saturation of the ferromagnet, unlike the solution with ZMP media in {(b)}.
Panels {(c)}, {(d)}, {(g)} and {(h)} show the 
\textbf{B} modulus along the vertical central axis for the panels on top of each.
} 
\end{figure}

\newpage

\begin{figure}[ht]
	\centering
	\hspace*{-0.75cm}
		\includegraphics[width=0.9\textwidth]{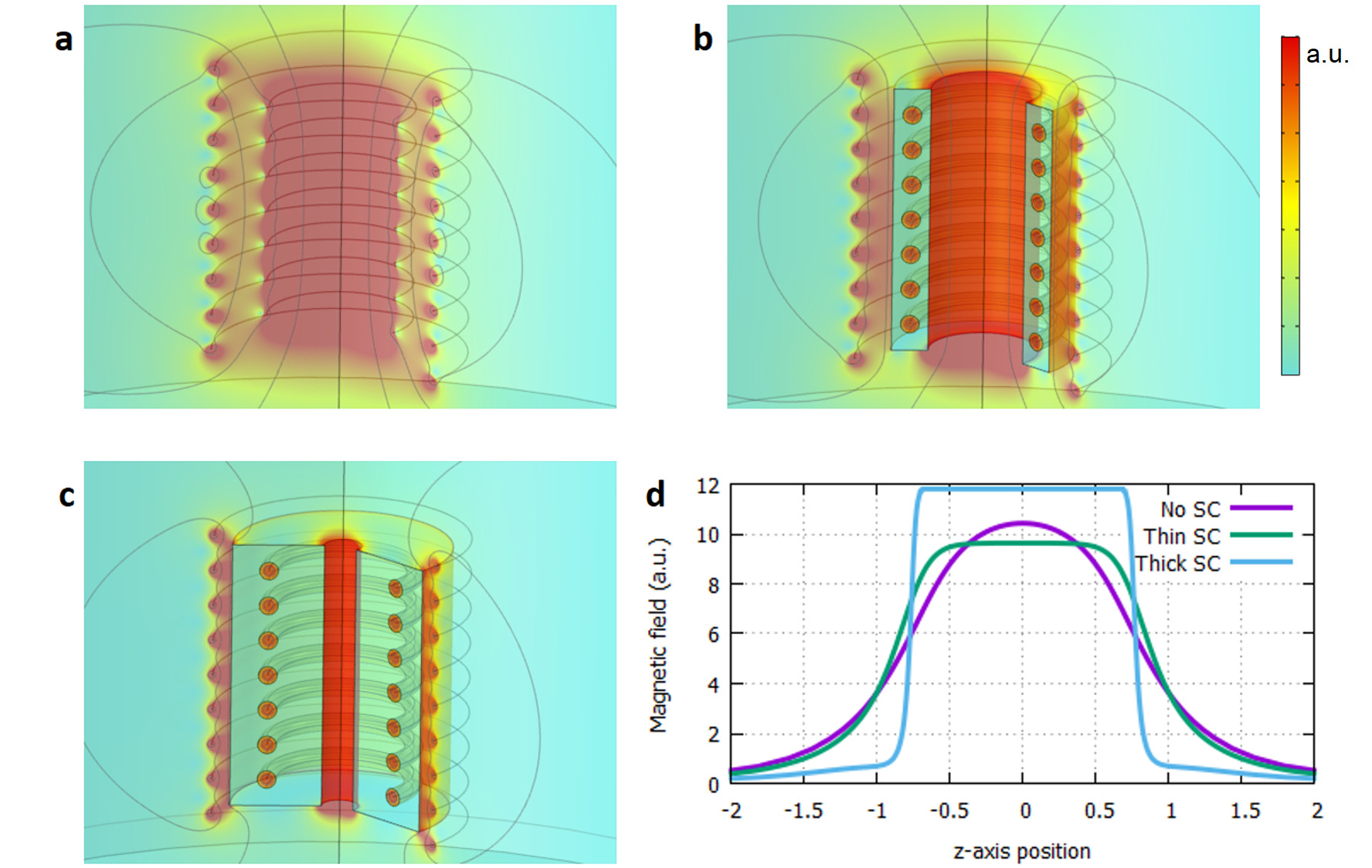}
	\caption{ 
	{(a)} Magnetic field {\bf B} modulus for a resistive-superconducting magnet, a sketch of the one in  \cite{larbalestier} that broke the highest magnetic-field record, consisting of an inner superconducting coil and a normal conducting outer one. {(b)} When enclosing the superconducting coil with ZMP media, e. g. a HTS bulk, the field in the core remains similar, while the force in the wires is eliminated. By changing the shape of the ZMP, for example, by reducing the gap,    
{(c)}, field can be further concentrated, made more homogeneous, or in general modified as desired. {(d)}  {\bf B} profiles at the vertical axis of the coils for cases {(a-c)}.
}	
\end{figure}

\clearpage

\newpage

\vskip 2truecm

\parindent=0pt

\section*{\Large Supplemental Material for 'Shaping Magnetic Fields with Zero-Magnetic-Permeability Media'}





\vskip 4truecm

{\parindent=2truecm

{\bf 
Contents:  

\medskip

- Methods.

- Supplementary Figures S1-S6.

- Description of Supplementary Video SV1.

}

}







\newpage

{\Large\bf Methods}

\vskip 1truecm


\subsection*{{Numerical simulations}}

Numerical simulations were performed using the program COMSOL Multiphysics 5.6 with the magnetic fields (mf) physics interface which solves the equations $\nabla\times\textbf{H}=\textbf{J}$, $\textbf{B}=\nabla \times \textbf{A}$ and $\textbf{J}=\sigma\textbf{E}$ with boundary conditions $\textbf{n}\times \textbf{A}=0$. The simulations were performed in the stationary state, using the 2D axisymmetric and 2D spatial dimensions, depending on the symmetry of each case. The simulation mesh selected had a maximum element size of 0.01m, minimum element size of 0.003m, a maximum element growth rate of 1.3, curvature factor of 0.3 and a resolution of narrow regions of 1.  The total calculation space was set to a 20m side length square with the relevant region of study of around 1m. 

In the case of Figure S6 the physics interface used is the Rotating Machinery, Magnetic (rmm) with the time dependent state and using 2D spatial dimensions.

\subsection*{{Magnetic field measurement}}

The magnetic field was measured using the Hall probe Arepoc model HPP-NP. This probe was connected to a source of DC current (Agilent 6654A  system DC power supply 60V, 9A) and a nanovoltmeter (Keithley 2182A) which displayed the potential difference across the probe. Each measurement of the magnetic field {\bf B} at a given position was made by collecting 4 different records of the potential difference across the probe, for 4 different currents (10mA, 17mA, 23mA and 30mA). With these values of voltage and current a linear regression was performed. The value of the obtained slope was used to determine the magnetic field, using a calibration line previously done by using the theoretical Ampere's law.


\subsection*{ {Measurement of the magnetic field of a current wire inside a superconducting tube}}

A superconducting hollow cylinder of Bi- 2223 with an outer radius of 1.15cm and an inner one of 1cm, and a length of 7.19cm was placed in a liquid nitrogen vessel. A current wire passing through the superconductor hole was connected to a current source KEPCO BOP 50-8M. The Arepoc Hall probe was placed at a fixed position
outside the superconducting cylinder (see the sketch in fig. 3(a)). The current wire was moved within the superconductor hole in the direction away and towards the probe.
The same measurements were repeated without the SC, for comparison. 

\subsection*{{Measurement of the magnetic field of currents in a two-holed superconducting cylinder}}

The double hollowed cylindrical SC is composed of 10 separate cylindrical pieces of YBa$_2$Cu$_3$O$_{7-x}$ each with height 1cm, radius 1.375cm and two inner holes with radii 0.5cm and 0.25cm, respectively, stacked on top of each other so that they form a cylinder with two perfectly aligned holes (see Supplementary Fig. S5). Two current wires were placed one at each hole and connected to DC current sources KEPCO BOP 50-8M and KEPCO BOP 100-4M. The Hall probe was placed outside the superconductor at a fixed position (see the sketch in Fig. 5a). The superconductor, the probe and the wires were all immersed in a vessel of liquid nitrogen. 
A current of +3A was applied to one of the cables and the current through the other wire was changed between +3A and -7A in steps of 1A, recording the magnetic field outside the superconductor with the probe.  The same experiments were carried out without the superconductor, only with the wires, for comparison. 


Using the same superconducting two-holed cylinder, a copper wire connected to a DC current source KEPCO BOP 50-8M was passed through the hole of radius 0.25cm and the probe was inserted in the middle of the 0.5cm hole in the centre of the cylinder (see the sketch in  Fig. 6b). The whole setup was immersed in liquid nitrogen.
The magnetic field in the probe position was measured for different values of current in the wire. The experiment was repeated without the superconductor, for comparison.

\newpage

\setcounter{figure}{0}
\renewcommand{\thefigure}{{S}\arabic{figure}}

{\Large\bf Supplementary Figures}


\vskip 1truecm

\begin{figure}[ht]
\centering
\includegraphics[scale=0.35]{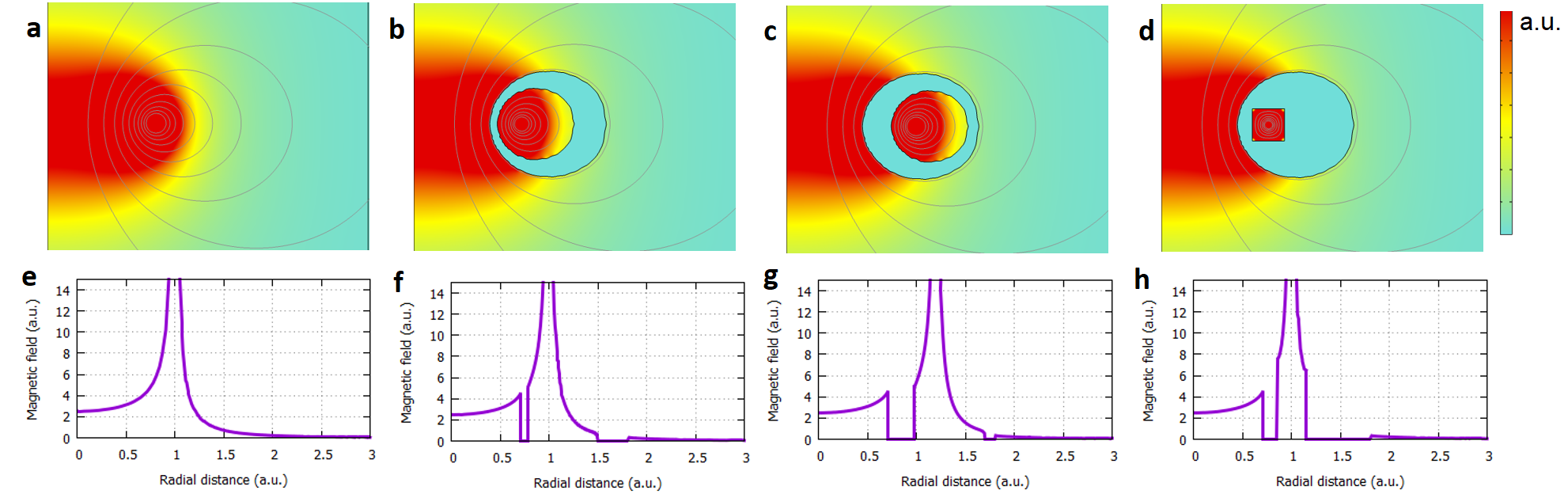}
\caption{
{(a-d)}
Magnetic field {\bf B}
modulus for {(a)} a bare current loop, and {(b-d)} differently shaped linear ZMP media surrounding a
current loop. The planes shown in the panels have rotational symmetry with respect to a central axis (the left edge of the images).
{(e-h)} The respective \textbf{B} modulus as a function of distance to the axis along the loop plane. 
The magnetic field of a current loop, {(a)}, when surrounded by a ZMP material with surfaces parallel to the {\bf B} field lines, as in {(b)} and {(c)}, is maintained unchanged inside and outside this ZMP material (compare {(f)} and {(g)} with {(e)}). When surfaces are not parallel to the field lines of the system, as in {(d)}, the magnetic field created by the current loop gets distorted with respect to that of the bare loop (compare 
{(h)} with {(e)}). 
} 
\label{reflectit}
\end{figure}
\newpage

\begin{figure}[ht]
\centering
\includegraphics[scale=0.45]{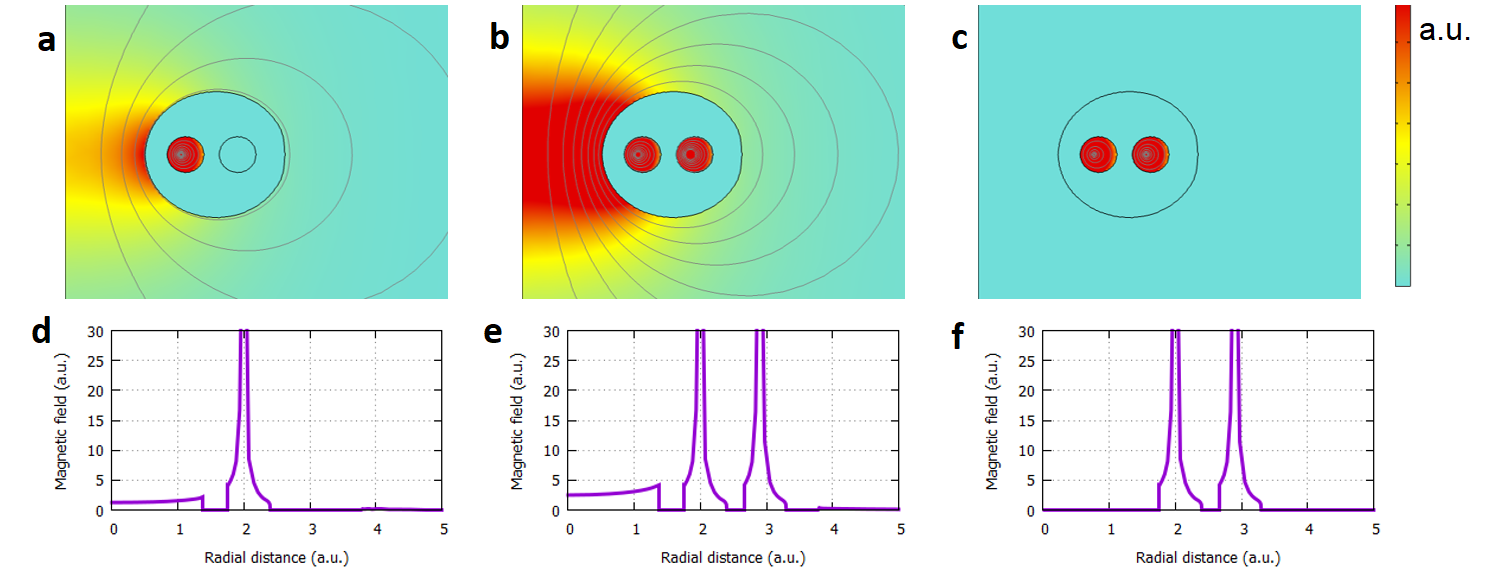}
\caption{
(a) Magnetic field {\bf B}
modulus for a 
current loop inside one of the two holes of a toroidal ZMP material, {(b)} two loops with the same current in separate holes of a toroidal ZMP material, {(c)} two loops with opposite currents in separate holes of a toroidal ZMP material. The planes shown in the panels have rotational symmetry with respect to a central axis (the left edge of the images).
{(d-f)}
The respective \textbf{B} modulus as a function of distance to the axis along the loop plane. 
The magnetic field outside the ZMP material always corresponds to the addition of the fields created by the net current passing through both loops. In the case of {(b)}, the field in the exterior corresponds to that of a centered loop with double current than that in {(a)}, and in {(c)} the exterior field is zero, as both current loops cancel each other out.
} 
\end{figure}

\newpage

\begin{figure}[ht]
\centering
\includegraphics[scale=0.3]{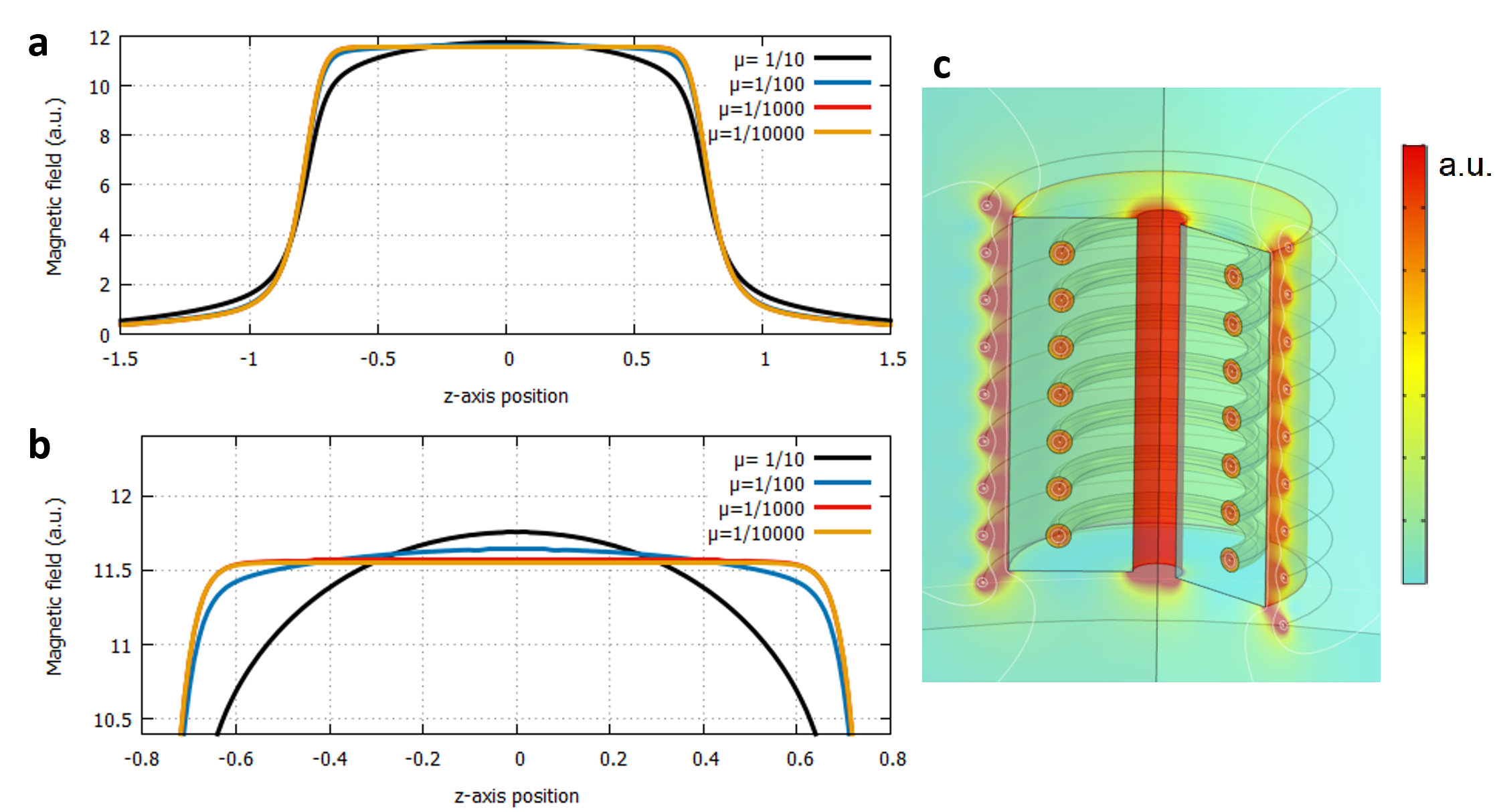}
\caption{(a) Magnetic field \textbf{B} (in arbitrary units) at the $z$-axis for the arrangement of coils and ZMP material shown in (c), for several values of the permeability $\mu$ of the ZMP material surrounding the coil. (b) Zoom of the graph (a). (c) Arrangement of coils and ZMP material to calculate the magnetic field at the z-axis (configuration like that in fig. 8c).
 } 
\label{reflectit}
\end{figure}

\newpage

\begin{figure}[ht]
\centering
\includegraphics[scale=0.3]{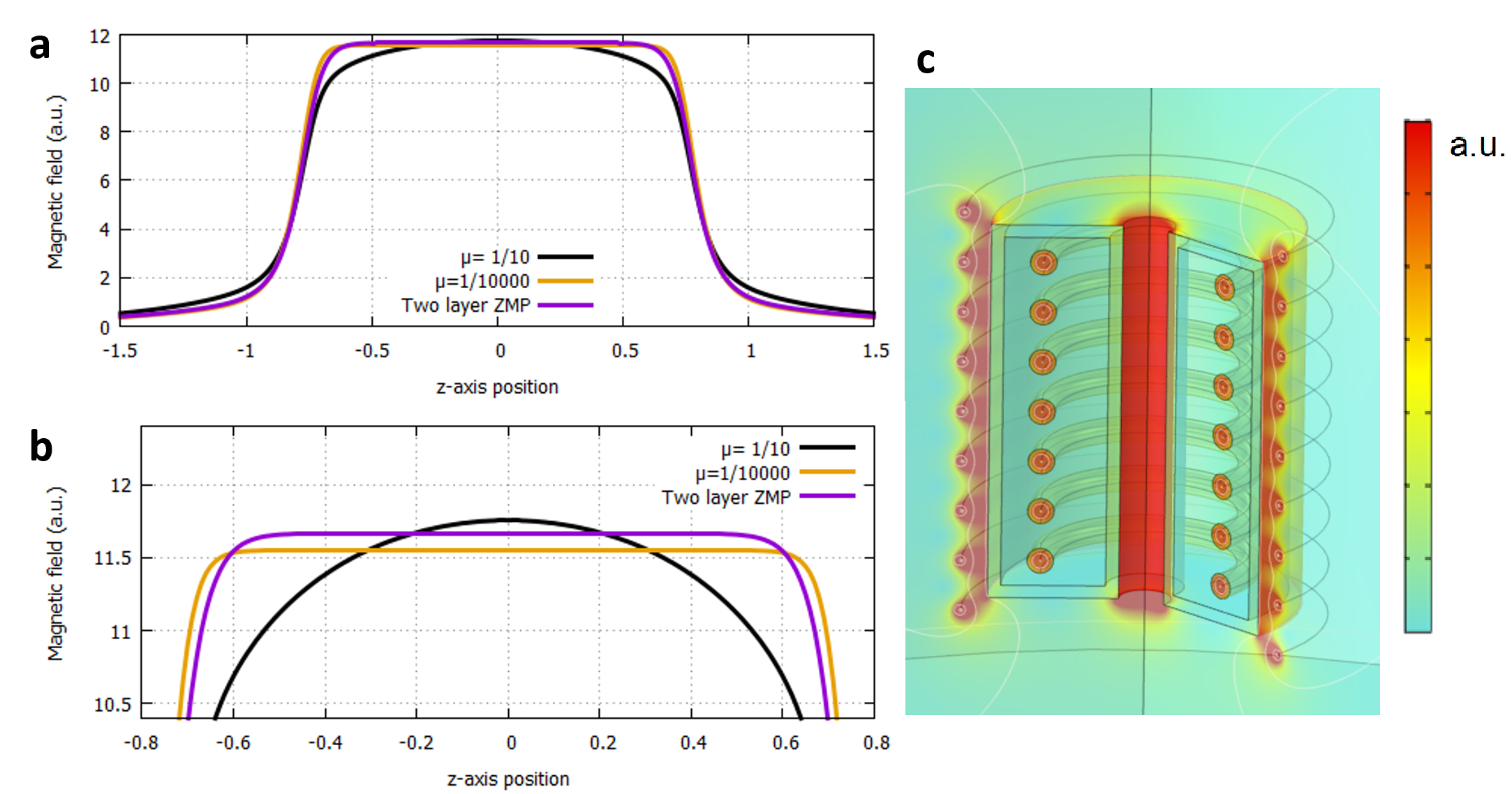}
\caption{
(a) Magnetic field \textbf{B} (in arbitrary units) at the $z$-axis for the arrangement of coils and ZMP material shown in (c), for a homogeneous high and low value of $\mu$, 1/10000 (yellow) and 1/10 (black), respectively, and a hybrid system of both, the lower $\mu$ at the surface and the higher at the center of the ZMP material (purple). (b) Zoom of the graph (a). (c) Arrangement of coils and ZMP material used to calculate \textbf{B} at the z-axes. For the hybrid-permeability case, the boundary division between both values of $\mu$ is represented as a black line.   } 
\end{figure}

\newpage

\begin{figure}[ht]
\centering
\includegraphics[scale=0.4]{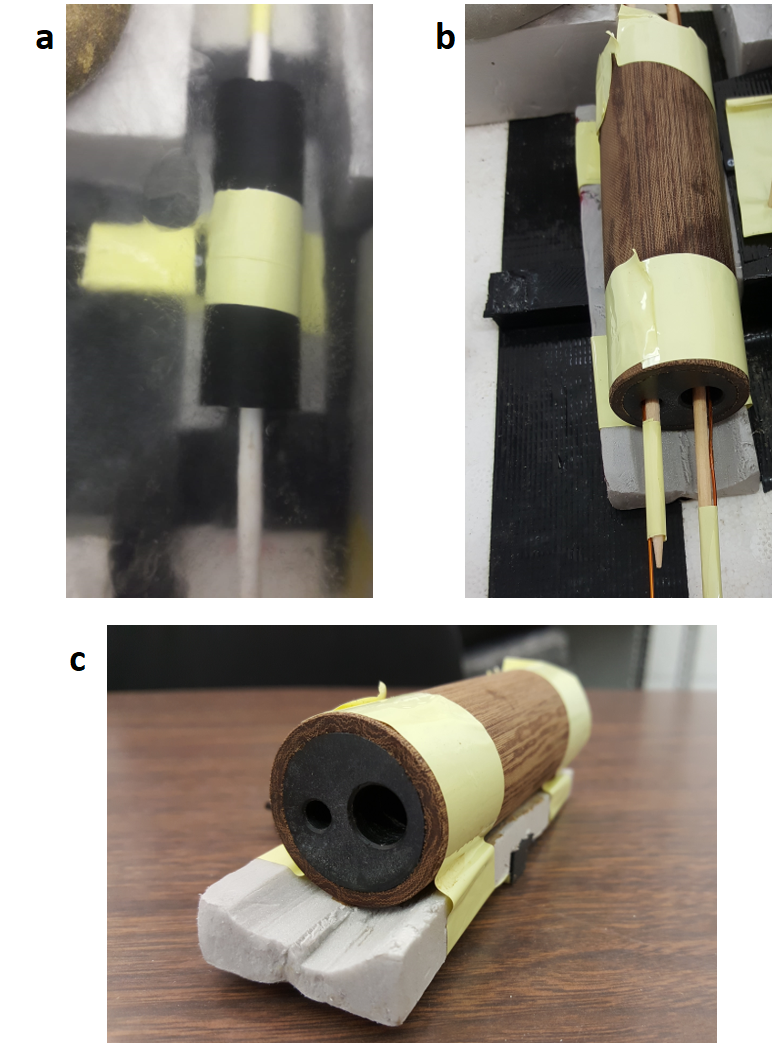}
\caption{
(a) Setup for the experiments of fig. 3 of the main text, when the Bi$_2$Sr$_2$Ca$2$Cu$_3$O$_{10-x}$ superconducting tube, the Hall probe (in the left) and the wire are all submerged in liquid nitrogen. {(b)} Setup of the experiment in fig. 5 of the main text, with the YBa$_2$Cu$_3$O$_{7-x}$ superconducting tube with two holes. In the case of the experiment in fig. 6 of the main text, the same setup as in {(b)} was used, with the cable shown on the right being replaced by the probe, which was located at the center of the cylinder. {(c)} View of the superconducting tube with two holes made of YBa$_2$Cu$_3$O$_{7-x}$ high-temperature superconductor, made of 10 adjoining pieces, with a resin covering. These images illustrate the feasibility of applying our ideas in practice, including the design of specific shapes for the high-temperature superconductor [e. g. the holes in the body shown in {(c)}].} 
\end{figure}

\newpage

\begin{figure}[ht]
\centering
\includegraphics[scale=0.6]{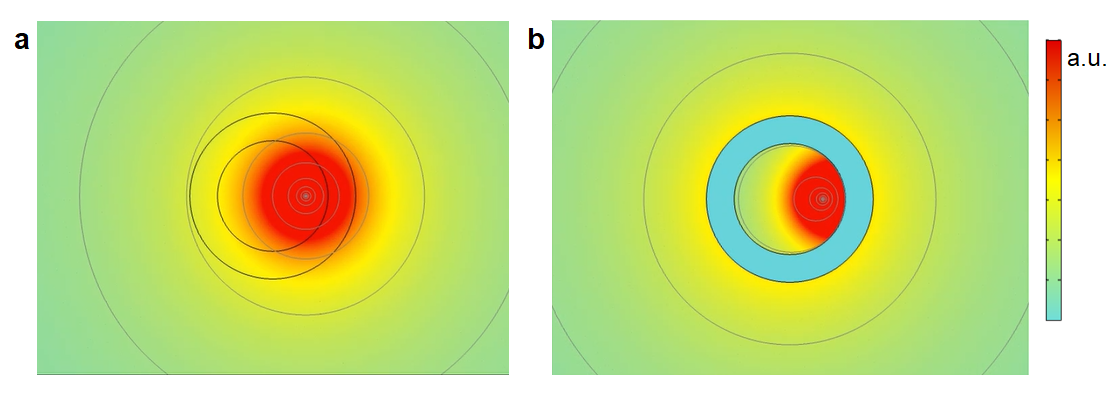}
\caption{Snapshots of the animation of the magnetic field {\bf B} around a current wire when it is rotating at a frequency of 1Hz, for the cases: (a) bare wire, and (b)
the current wire is inside a cylindrical hollow ZMP material. } 
\label{reflectit}
\end{figure}

\bigskip

{\Large\bf Description of Supplementary Video SV1:}

Animation of the magnetic field {\bf B} around a current wire when it is rotating at a frequency of 1Hz, for the cases: (left) bare wire, and (right)
the current wire is inside a cylindrical hollow ZMP material.

\end{document}